\documentclass[conference]{IEEEtran}
\IEEEoverridecommandlockouts

\usepackage{cite}
\usepackage{amsmath,amssymb,amsfonts,mathtools}
\usepackage{graphicx}
\usepackage{siunitx}
\usepackage{epstopdf}
\usepackage{textcomp}
\usepackage[hyphens]{url}

\usepackage[breaklinks,hidelinks]{hyperref}
\usepackage{breakurl}
\usepackage{xcolor}
\def\BibTeX{{\rm B\kern-.05em{\sc i\kern-.025em b}\kern-.08em
    T\kern-.1667em\lower.7ex\hbox{E}\kern-.125emX}}

\begin{document}

\title{Impact of COVID-19 Pandemic on Energy Demand in Estonia \\
\thanks{The work was partly supported by the Estonian Research Council grants COVSG22 and PRG1463. Authors also would like to thank Mr. L. Kadakas for valuable inputs to this manuscript, and Elektrilevi O{\"U} for provided data.}
}

\author{\IEEEauthorblockN{Juri~Belikov}
\IEEEauthorblockA{\textit{Department of Software Science,} \\
\textit{Tallinn University of Technology,}\\
Akadeemia tee 15a, 12618 Tallinn, Estonia \\
juri.belikov@taltech.ee}
\and
\IEEEauthorblockN{Jaan~Kalda}
\IEEEauthorblockA{\textit{Department of Cybernetics,} \\
\textit{Tallinn University of Technology,}\\
Akadeemia tee 21, 12618 Tallinn, Estonia \\
kalda@ioc.ee}
}

\maketitle

\begin{abstract}
The COVID-19 pandemic triggered a question of how to measure and evaluate adequacy of the applied restrictions. Available studies propose various methods mainly grouped to statistical and machine learning techniques. The current paper joins this line of research by introducing a simple-yet-accurate linear regression model which eliminates effects of weekly cycle, available daylight, temperature, and wind from the electricity consumption data. The model is validated using real data and enables the qualitative analysis of economical impact.
\end{abstract}

\begin{IEEEkeywords}
COVID-19, Energy modeling, Load forecast
\end{IEEEkeywords}

\section{Introduction}
On March 11, 2020, the World Health Organisation (WHO) declared the novel coronavirus (COVID-19) to be a global outbreak \cite{Cucinotta2020}. A day after, local transmission of the virus was found in Estonia \cite{ERR2020a}, followed by the state of emergency announced in Estonia, commonly known as a ‘lockdown’ \cite{ERR2020b}.

One consequence of the COVID-19 pandemic was a significant change in generation and consumption of energy mix worldwide. Many countries have reported significant decrease \cite{Navon2021}. Therefore, it becomes important to study and analyze the impact COVID-19 on electricity demand patterns, since errors in the existing models can lead to substantial monetary losses in both public and private sectors. Seasonal and meteorological changes have affect electricity demand and need to be considered when developing new models \cite{Salisu2016}. Hence, for better understanding it is imperative to eliminate the weather footprint from the data. It is also important to consider private and business consumers separately, as their energy behaviours depend on the applied measures. Recent attempt to model pandemic effects can be found in \cite{Wang2020}. However, general future pandemic-proof solutions, being useful for grid operators, are still to be developed.

In this light, work \cite{Alasali2021} reported investigations on the impact of COVID-19 pandemic on energy consumption, where a separate rolling stochastic Auto Regressive Integrated Moving Average with Exogenous model was created. The work analyzes impact the pandemic had on Jordan’s power grid, including a reduction in average energy consumption of $40\%$ in 2020 compared to 2019. Paper \cite{Garcia2021} shows that the full lockdown in Italy caused a $25\%$ decrease in energy consumption, $13.5\%$ in Spain, and $14\%$ in the province of Ontario, Canada. In the same study, residential and non-residential clients are considered from a town in Spain and shown that their behaviours are different, where residential clients show an increase of $13.1\%$ in consumption during lockdowns and non-residential clients had a reduction of $35.2\%$. Comparison of energy consumption data in the US from 2019 and 2020 citing probable weather differences is presented in \cite{Agdas2020}. The study focuses on Gainesville Regional Utilities (GRU) in Florida and applied a weather correction method. Overview of the recent results  and existing challenges can be found in \cite{Navon2021,Jiang2021}.

In this paper, we apply a 10-dimensional linear regression model which eliminates the effects of weekly cycle, available daylight, temperature, and wind from the electricity consumption time series, and brings the amplitude of unexplained fluctuations for a pre-COVID-19-period down to ca $1-2\%$. With the help of this model, the impact of the COVID-19 pandemics can be seen unhinderedly. Estonia is used as a case study, and depending on the county, the magnitude of the impact appears to range from nearly $0\%$  up to ca $30\%$. Another benefit of the proposed model is simplicity enabling near real-time decision, and suitability for evaluation of the applied measures and impact on the economy.

\section{Data and Methodology}

\textit{Weather dataset}: The weather dataset was obtained from a public weather station in Tallinn airport\footnote{Currently, the used publicly available and reliable source of information provides detailed data only for one city in Estonia.}, and contained rich variety of data. For our purposes, the necessary fields are timestamps, air temperature at two meter height from Earth’s surface, mean wind speed, and total cloud coverage. The retrieved data however contained some missing samples, which were filled in with approximations of what the values should have been by following the trend of nearby values. 

\textit{Electricity consumption dataset}: The electricity consumption data was provided by the local distribution system operator Elektrilevi O{\"U} to Tallinn University of Technology for study and research purposes under agreement JV-ARI-18/27941. The electricity consumption dataset is a large collection of daily electricity consumption measurements from approximately 10\,000 locations, covering the entire country. Data also differentiate business and private consumers. This was later aggregated together for each county and the type of consumer. The measurements had few outliers, which have been fixed by replacing them by averaging neighbour days.

\textit{Solar calculations}: Solar calculations, obtained from National Oceanic and Atmospheric Administration, were correct.

\subsection{Model description}
The daily electricity consumption of both business and private consumers fluctuates significantly, due to the weekly cycle, weather conditions, solar irradiance, etc. The typical magnitude of such fluctuations, for a county, or for the country as a whole, is typically around $20\%$. These fluctuations obstruct the effect of additional factors, such as the effect of COVID-19 pandemics. We applied the multiple linear regression method to compensate the fluctuations.

Specifically, let $c_i$ ($i=1,\ldots,N$) denote the data vector of electricity consumption (with index $i$ denoting the order number of the day), and $f_i^\mu$ ($\mu=1,\ldots, M$) denoting $M$ factors affecting consumption; in what follows, $M=10$. Then, the vector of residuals $r_i=c_i-\sum_\mu \alpha_\mu f_i^\mu$ of this $M$-dimensional least-square linear fit of the vector $c_i$ with the factors $f_i^\alpha$ is the vector of ``unexplained'' fluctuations, including the effect of pandemics. Here, $\alpha_\mu f_i^\mu$ represents the prediction vectors (listed below), and coefficients $\alpha_\mu$ are solutions of the linear set of equations
\begin{equation*}
\sum_i\alpha_\mu f_i^\mu f_i^\nu=\sum_ic_if_i^\nu,
\end{equation*}
where the first seven prediction vectors are the weekdays. It appears that the weekly fluctuation cycle cannot be reduced to a simple differentiation between business days and weekends. For instance, the business consumption on Mondays is typically smaller than on Tuesdays. So, we took each of the seven weekdays as a factor: $f_i^1=1$ if the $i$th day is Monday, and $f_i^1=0$ otherwise; $f_i^2=1$ if the $i$th day is Tuesday, and $f_i^2=0$ otherwise; etc. These seven factors are linearly dependant with a constant vector $f_i^0\equiv 1$. Therefore, the regression had to be done while forcing the constant term to be equal to zero. The other factors are as follows: the thermal effect $f_i^8=|T_i-\SI{20}{\celsius}|$, where $T_i$ denotes average temperature during the $i$th day; the modulus is taken, since we assume that electricity is consumed for heating when it is too cold outside, and for cooling when it is too warm. The vector $f_i^9$ reflects the solar radiation, hourly-weighted for each day, expressed in effective number of hours with available daylight, which approximately explains the following three effects: (a) lightning-related energy consumption is expected to depend on the weighted number of hours with available daylight; (b) solar radiation arriving through windows heats rooms reducing thereby heating-related expenses; (c) solar radiation is reducing the need for external energy supply of households with private solar panels. The current study is based on a single weather station, and calculated based on solar altitude angle and cloud cover. Using direct solar radiation measurements data, and averaging over an array of weather stations, is expected to reduce the noise to levels low enough to enable meaningful separation of the available light and solar radiation prediction vectors. Finally, the vector $f_i^{10}$ denotes heat losses\footnote{These heat losses are expected to be proportional to the dynamic wind pressure which is proportional to the squared wind speed.} due to wind measured as the average squared wind speed, multiplied by $f_i^8$.

With our $10$ factors, during the period preceding the pandemics, the magnitude of the normalized residuals $\rho_i=r_i/\langle c_i\rangle$, which represent the unexplained fluctuations, was brought down to around $1-2\%$. Here, the angular braces denote averaging over a one-month-long sliding window centered around the $i$th day (except after the onset of pandemics, in which case the averaging is performed over the last 30 days before the onset).

This regression model appears to be robust enough to allow a fairly long extrapolation: when the regression coefficients are found for a 30-day period, the obtained linear fit can be extrapolated up to five months, while keeping the residual consumption fluctuations below $2\%$. This observation allows us to obtain the regression coefficients based on the period prior to the onset of pandemics, and extrapolate the prediction vectors into the pandemics. However, during the pandemics, a small additional procedure proved to be necessary: with a strong overall decline in consumption (around $20\%$), the amplitude of the weekly consumption cycle appeared to have been decreased, as well. This would result in an overcompensation of this weekly cycle if the prediction vector were to be based on the preceding period with higher overall consumption. In order to avoid such over-compensation, the prediction vectors need to be multiplied with a factor corresponding to the global consumption decrease (for instance, with $20\%$ fall due to pandemics, the prediction vectors were multiplied by $0.8$).

The length of the training period, for finding the regression coefficients, was taken equal to 30 days due to the following considerations. We aimed to use as short as possible period, so as to avoid the effect of potential long-term trends in the consumption of business consumers, unrelated neither to our prediction vectors nor COVID-19. The 30-day period proved to be just sufficient, but this is only owing to the fact that the period used (February) is the period of rapid increase in the available daylight. Closer to solstice, the training period needs to be increased at least by a factor of two.

\section{Numeric results}
The proposed model was first used to calculate the normalised residuals for 2019 to validate the accuracy. The training period was 31 days from January 1 to January 31. The model showed good validation results for up to five months. Hence, for a general overview of the change in consumption, the prediction period was chosen from February~1 to June~30.

\begin{figure}[!t]
\centering
\includegraphics[width=0.9\columnwidth]{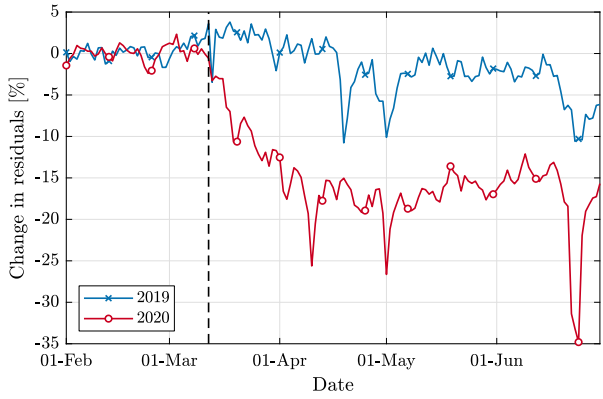}
\caption{The normalised residuals showing general trend in change of electricity consumption for business consumers. Dashed line denotes the moment of declaring the emergency state, March 12, 2020.}
\label{fig:totB}
\end{figure}

The normalised residuals for 2019 and 2020 are presented in Fig.~\ref{fig:totB}. There are several drops for both years on May 1, which is due to the May Day state holiday. A second drop happens during April 19 of 2019, and April 10 of 2020, which is due to the Good Friday state holiday. The third drop happens from June 22 to 24, which coincides with Victory Day and Midsummer Day. However, in 2019, these happened on weekends, therefore their drop was smaller.

\begin{figure}[!t]
\centering
\includegraphics[width=0.9\columnwidth]{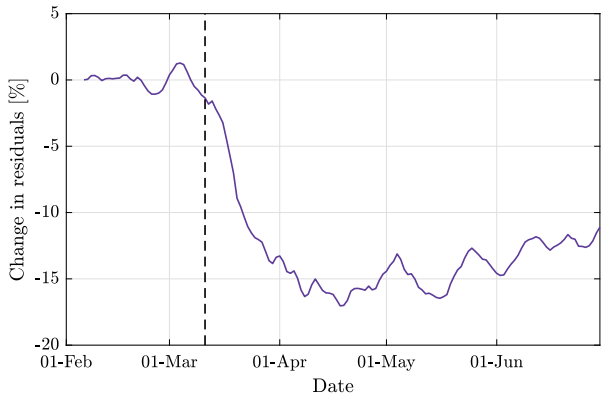}
\caption{Moving average difference between residuals of 2019 and 2020.}
\label{fig:totDiffB}
\end{figure}

For a better understanding of the general change in consumption, the residuals in 2020 are subtracted from the residuals in 2019, which is shown in Fig.~\ref{fig:totDiffB}. Since the year 2020 was a leap year, the residual for February 28 is recorded twice for the year 2019. The difference also accounts for movable state holidays, where the correct date and the surrounding dates are chosen for comparison instead. For example, the Good Friday was on April 19 of 2019 and April 10 of 2020, so the dates April 18 to 20 of 2019 were compared with the dates April 9 to 12 of 2020.  Conversely, April 9 to 12 of 2019 was compared with April 18 to 20 of 2020 to account for the missing dates. Figure~\ref{fig:totDiffB} shows how the difference reduces as restrictions are relaxed, gradually stabilising throughout the summer. There is still a general loss in consumption, which might be due to many still opting to stay at home, work from home, or generally travel less during the pandemic. To incorporate weekly cycles, a 7-days moving average is calculated, smoothing small spikes from Fig.~\ref{fig:totB}.

\subsection{Selected counties-based analysis}
In this section, we present analysis for several selected counties for either business or private consumers, based on population density (Harju), industrialization level (Ida-Viru), or severity of applied restrictions (Saare).

\subsubsection{Harju county}
\begin{figure}[!t]
\centering
\includegraphics[width=0.9\columnwidth]{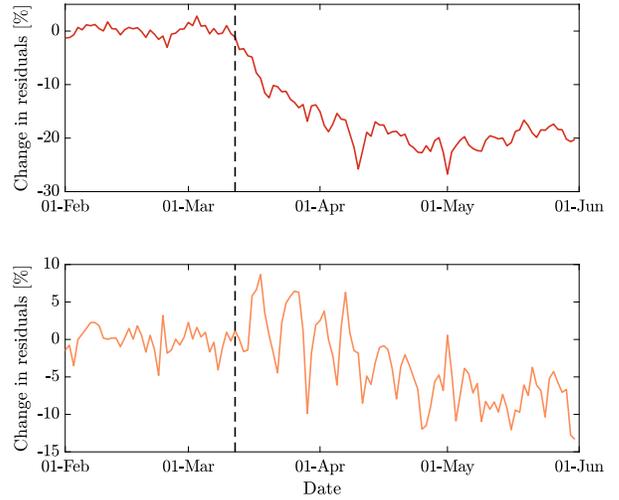}
\caption{Change in residuals for business (top) and private (bottom) consumers in Harju county, without weekly cycle normalisation.}
\label{fig:harjuBP}
\end{figure}

The change in consumption for business clients is presented in top plot of Fig.~\ref{fig:harjuBP}, revealing almost immediate drop in consumption a week after the lockdown was announced. Within three weeks, the consumption drops by almost $15\%$ and continued decreasing to about $-20\%$ until April, with the lowest change in consumption excluding holidays recorded on April 25 as $-22.7\%$.

The bottom plot in Fig.~\ref{fig:harjuBP} depicts Harju county's private consumer change. The weekly cycle is more visible throughout with earlier weeks having an amplitude of roughly $12\%$. The fluctuations stabilize towards the end of the lockdown with an amplitude of $7\%$. Unlike a constant decrease observed for business consumers, the lockdown amplified the weekly cycles for private consumers.

\subsubsection{Ida-Viru county}
\begin{figure}[!t]
\centering
\includegraphics[width=0.9\columnwidth]{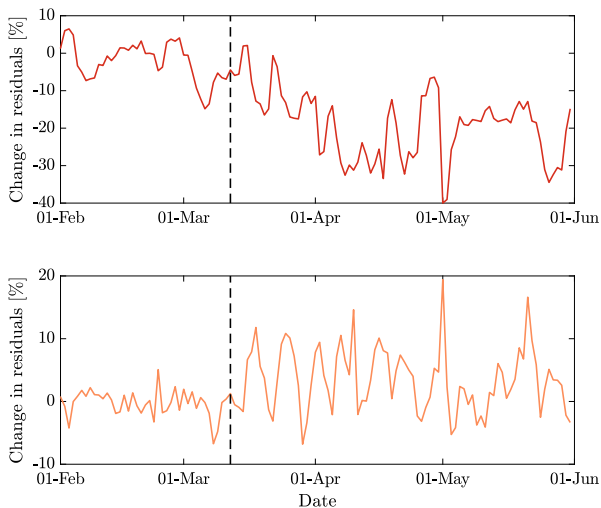}
\caption{Change in residuals for business (top) and private (bottom) consumers in Ida-Viru county.}
\label{fig:iviBP}
\end{figure}

The consumption change for Ida-Viru business consumers is shown in top plot Fig.~\ref{fig:iviBP}, where the weekly cycle is more distinct. There are larger fluctuations, but a general loss is shown, with the earlier weeks fluctuating between $-5\%$ and $-15\%$. During the month of April, the consumption stays at a near $-30\%$ low. There is a large drop on May Day, but the surrounding days are below $-20\%$. During most of May, the consumption stays near $-20\%$, with a sudden drop around May 26th to the lowest of the entire lockdown period, $-34.5\%$.

The bottom plot in Fig.~\ref{fig:iviBP} shows the consumption change for private consumers in Ida-Viru county, with a noticeable increase in general during the lockdown period. The largest spike in consumption is on May Day, but outside of state holidays, there are similarly large spikes on other weeks, with the highest being on May 21, with an increase of $17.3\%$. Mostly, electricity consumption rose by about $10\%$ for private consumers with a slight decrease towards the end of lockdown.

\subsubsection{Saare county}
\begin{figure}[!t]
\centering
\includegraphics[width=0.9\columnwidth]{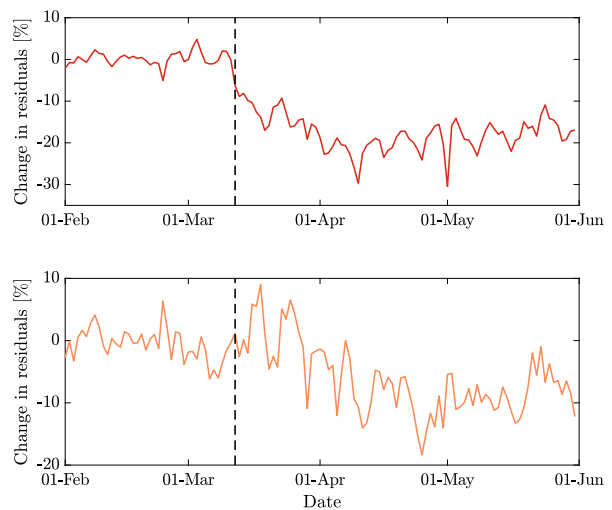}
\caption{Change in residuals for business (top) and private (bottom) consumers in Saare county.}
\label{fig:saareBP}
\end{figure}

The lockdown has a clear impact for Saare county business consumers for the duration of March, as shown in top plot in Fig.~\ref{fig:saareBP}. The demand stabilized in April and showed a small uptrend throughout May. The first week of the lockdown thus showed the most drastic change, a change from nearly $0$ to about $-17\%$ in demand. The weekly fluctuations, however, were smaller for throughout three months.

Saare county was first and hardest hit by the COVID-19 virus during the spring. Stay-at-home orders were issued, and all western islands were quarantined. There was still a small downtrend in demand for private consumers, though initially the demand increased to nearly $10\%$, immediately after restricting movements. The increase in demand continued for the rest of March, with high fluctuations in consumption during early April, see bottom plot in Fig.~\ref{fig:saareBP}. The demand continued to lower and stabilized by early May, staying near $-10\%$, and increasing again at the end of May.

\section{Analysis and Discussion}
In this section, results are analysed and compared with statistical data provided by the Ministry of Finance and Statistics, Estonia. The goal is to understand whether the government restrictions were efficient or too harsh for any separate county.

\subsection{Comparison with data from Statistics Estonia}
The monthly turnover of enterprises chart in year 2020 shows a loss of about $-17\%$ between March and April. During May, the turnover increases substantially, but does not reach pre-pandemic levels, stagnating in June and dropping slightly again in July \cite{SE2020a}.

Comparison of years 2020 and 2019 reveals a $-22.6\%$ decrease for April. The year 2019 had a small increase during March and April, and a dip in May. The energy consumption in Estonia followed a similar trend by dropping around $16\%$ during April and gradually recovering from there, with small jumps in May. To reduce revenue losses, many companies had to lay off their staff or close their doors entirely---the most severely affected were accommodation and food service industries \cite{SE2020b}. This could have also led to an increase in consumption for private consumers, as they would have stayed at home. However, the government put aside relief efforts for most affected enterprises for the lockdown period and larger layoffs happened instead during June-July, when the relief benefits likely would have been ended \cite{VV2020}.

\subsection{Comparison with data from Ministry of Finance}
Statistics from Ministry of Finance \cite{RM2020} showed a large decrease in the monthly revenue change. Thus, March had a $-18.5\%$ decrease, April continuing the descent to $-20.4\%$; May bounced to $-7.4\%$.

Estonian Institute of Economic Research reported several problems the pandemic has had on enterprises, with $68\%$ of enterprises reporting a lowered demand in production, $42\%$ reporting a partial lack of employees (due to illnesses or quarantines), $41\%$ finding it necessary to send employees on paid leave, and $25\%$ of enterprises making employees redundant for the month of March. This correlates with the patterns for consumption change during lockdown, as lower demand for production also directly affects the need for energy consumption, and lack of employees both lowering to a smaller extent the business consumer energy consumption, while also increasing the private consumer consumption.

\subsection{Patterns in economic losses and energy consumption}
The data from Statistics Estonia and Ministry of Finance both follow a similar pattern to the general decline in energy consumption calculated by the proposed regression model. Moreover, the data also suggests that counties with larger industries or higher density in population were hit harder due to problems caused by the pandemic, leading to a harsher drop in energy consumption.

Counties with a larger share of private consumers however were not hit as hard, or even reported an increase in energy consumption, as indicated by southern counties, mostly explained by the lockdown and stay-at-home orders, or layoffs from companies. At the same time, the increased consumption from private consumers caused the weekly cycles for private consumers to amplify.

\subsection{Analysis of county results and measures taken}
Hiiu and Saare counties are not very large compared to Harju, Ida-Viru or L{\"a}{\"a}ne-Viru counties. However, they reported similar drops in proportion for business consumers. While Harju and Ida-Viru counties showed an increase in private consumer consumption, two islands still reported a decrease. This can be explained by the additional measures introduced to prevent the initial spread of the COVID-19 virus in Saaremaa and other western islands.

While it is difficult to truly verify whether the government took adequate measures or not to avoid spread of the virus, it is possible to see through energy consumption which counties were affected the most. It is expected for larger industrial counties to take the largest hits. Therefore, for smaller counties such as Hiiu and Saare to report similar numbers could have been too harsh for business consumers, especially if the economic losses follow a similar trend for the counties as well.

When the western islands went under additional restrictions to stop spread of the virus, this further lowered energy consumption. While some southern counties also reported outbreaks, Saare county was the most affected in spread initially, causing people to stay at home even more than in other counties. This with additional restrictions could have led to the difference in private and business consumer declines in consumption, whereas the southern counties reported a growth.

\begin{figure}[!t]
\centering
\includegraphics[width=0.9\columnwidth]{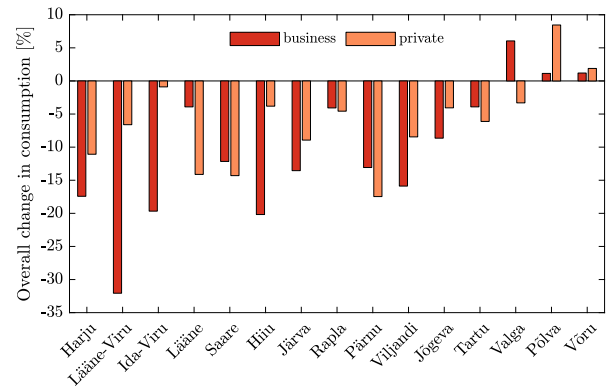}
\caption{Overall change in consumption for private and business consumers from March 3 to May 31.}
\label{fig:barsBP}
\end{figure}

Figure~\ref{fig:barsBP} shows the overall change in consumption for business and private consumers for each county. The most affected counties were P{\"a}rnu, Saare, and L{\"a}{\"a}ne in case of private consumers. In case of business consumers the most affected counties were L{\"a}{\"a}ne-Viru, followed by Harju, Hiiu, and Ida-Viru. Smaller counties (e.g., L{\"a}{\"a}ne) in population density or industries might show drastic changes due to weaker data, where the large change might be due to some companies starting or ending work.

Similarly, southern counties and counties with low density might also show larger numbers in growth or smaller declines. Many people have their private residences or summer houses in counties with sparse population and might have travelled there to stay distant from the crowdy areas when lockdown was announced. This could have been one of the reasons why these counties were less affected by the pandemic.

\bibliographystyle{IEEEtran}
\bibliography{main}

\begin{thebibliography}{10}
\providecommand{\url}[1]{#1}
\csname url@rmstyle\endcsname
\providecommand{\newblock}{\relax}
\providecommand{\bibinfo}[2]{#2}
\providecommand\BIBentrySTDinterwordspacing{\spaceskip=0pt\relax}
\providecommand\BIBentryALTinterwordstretchfactor{4}
\providecommand\BIBentryALTinterwordspacing{\spaceskip=\fontdimen2\font plus
\BIBentryALTinterwordstretchfactor\fontdimen3\font minus
  \fontdimen4\font\relax}
\providecommand\BIBforeignlanguage[2]{{%
\expandafter\ifx\csname l@#1\endcsname\relax
\typeout{** WARNING: IEEEtran.bst: No hyphenation pattern has been}%
\typeout{** loaded for the language `#1'. Using the pattern for}%
\typeout{** the default language instead.}%
\else
\language=\csname l@#1\endcsname
\fi
#2}}

\bibitem{Cucinotta2020}
D.~Cucinotta and M.~Vanelli, ``{WHO} declares {COVID-19} a pandemic,''
  \emph{Acta Bio Medica Atenei Parmensis}, vol.~91, pp. 157--160, 2020.

\bibitem{ERR2020a}
{ERR}, ``Terviseamet: {E}estis on kinnitatud 27 koroonajuhtu ja kohalik
  levik,'' 2020, [Online] Available
  \url{https://www.err.ee/1063204/terviseamet-eestis-on-kinnitatud-27-koroonajuhtu-ja-kohalik-levik},
  Accessed November 20, 2021.

\bibitem{ERR2020b}
{ERR}, ``Valitsus kuulutas välja eriolukorra,'' 2020, [Online] Available
  \url{https://www.err.ee/1063213/valitsus-kuulutas-valja-eriolukorra},
  Accessed November 20, 2021.

\bibitem{Navon2021}
A.~Navon, R.~Machlev, D.~Carmon, A.~E. Onile, J.~Belikov, and Y.~Levron,
  ``Effects of the {COVID}-19 pandemic on energy systems and electric power
  grids---{A} review of the challenges ahead,'' \emph{Energies}, vol.~14,
  no.~4, p. 1056, 2021.

\bibitem{Salisu2016}
A.~A. Salisu and T.~O. Ayinde, ``Modeling energy demand: {S}ome emerging
  issues,'' \emph{Renew. Sust. Energ. Rev.}, vol.~54, pp. 1470--1480, 2016.

\bibitem{Wang2020}
B.~Wang, Z.~Yang, J.~Xuan, and K.~Jiao, ``Crises and opportunities in terms of
  energy and {AI} technologies during the {COVID}-19 pandemic,'' \emph{Energy
  and {AI}}, vol.~1, p. 100013, 2020.

\bibitem{Alasali2021}
F.~Alasali, K.~Nusair, L.~Alhmoud, and E.~Zarour, ``Impact of the {COVID}-19
  pandemic on electricity demand and load forecasting,'' \emph{Sustainability},
  vol.~13, no.~3, p. 1435, 2021.

\bibitem{Garcia2021}
S.~Garc{\'{\i}}a, A.~Parejo, E.~Personal, J.~I. Guerrero, F.~Biscarri, and
  C.~Le{\'{o}}n, ``A retrospective analysis of the impact of the {COVID}-19
  restrictions on energy consumption at a disaggregated level,'' \emph{Appl.
  Energ.}, vol. 287, p. 116547, 2021.

\bibitem{Agdas2020}
D.~Agdas and P.~Barooah, ``Impact of the {COVID}-19 pandemic on the {U.S.}
  electricity demand and supply: {A}n early view from data,'' \emph{{IEEE}
  Access}, vol.~8, pp. 151\,523--151\,534, 2020.

\bibitem{Jiang2021}
P.~Jiang, Y.~V. Fan, and J.~J. Kleme{\v s}, ``Impacts of {COVID-19} on energy
  demand and consumption: {C}hallenges, lessons and emerging opportunities,''
  \emph{Appl. Energ.}, vol. 285, p. 116441, 2021.

\bibitem{SE2020a}
{Statistica Estonia}, ``Short-term statistics of enterprises,'' 2020, [Online]
  Available
  \url{https://www.stat.ee/en/find-statistics/covid-19-impact-estonia/short-term-statistics-enterprises},
  Accessed November 20, 2021.

\bibitem{SE2020b}
{Statistica Estonia}, ``Short-term labour market statistics,'' 2020, [Online]
  Available
  \url{https://www.stat.ee/en/find-statistics/covid-19-impact-estonia/short-term-labour-market-statistics},
  Accessed November 20, 2021.

\bibitem{VV2020}
{Vabariigi Valitsus}, ``Valitsus kiitis heaks covid-19 piirangutest enim
  m{\~o}jutatud ettev{\~o}tete toetamise,'' 2020, [Online] Available
  \url{https://www.valitsus.ee/uudised/valitsus-kiitis-heaks-covid-19-piirangutest-enim-mojutatud-ettevotete-toetamise},
  Accessed November 20, 2021.

\bibitem{RM2020}
{Rahandusministeerium}, ``Majandusn{\"a}itajad,'' 2020, [Online] Available
  \url{https://www.rahandusministeerium.ee/et/majandusnaitajad}, Accessed
  November 20, 2021.

\end{thebibliography}

\end{document}